\documentstyle[pra,epsf,aps,twocolumn]{revtex}

\def\cm#1{}

\begin{document}
\title{{
Perturbative Calculation of Multi-Loop Feynman Diagrams.\\
New Type of Expansions for Critical Exponents
}}
\author{Hagen Kleinert%
 \thanks{Email: kleinert@physik.fu-berlin.de \hfil \newline URL:
http://www.physik.fu-berlin.de/\~{}kleinert \hfil
}}
\address{Institut f\"ur Theoretische Physik,\\
Freie Universit\"at Berlin, Arnimallee 14,
14195 Berlin, Germany}
\maketitle
\begin{abstract}
We show that the calculation of $L$-loop Feynman integrals
in $D$ dimensions can be reduced to a series of matrix multiplications
in $D\times L$ dimensions. This leads to a new type of expansions
for the critical exponents in three dimensions in which all coefficients
are calculable exactly.
\end{abstract}

%
~\\
\noindent
{\bf 1.}
Recently, an efficient method has been developed
for constructing all Feynman  diagrams of a quantum field theory
recursively to arbitrary order in perturbation theory \cite{rec}.
The construction has also been automatized
by computer algebra \cite{rec}, so that one can now easily list
all diagrams to any desired order,
limited only by computer time.

The only open problem is an efficient evaluation
of the associated Feynman integrals.
The highest loop number so far has been treated in $\phi^4$-theory
where renormalization constants
of field normalization and mass are now known
numerically up to
seven loops  \cite{MN}.

For a reliable calculation of all critical exponents
with an accuracy which has become recently available
in satellite experiments on superfluid helium
\cite{Lipa} it is desirable to
know the perturbation expansions to even higher order \cite{seval,seven}.
In this note we would like to propose
a perturbative evaluation
of many-loop Feynman integrals
which reduces the integrations
over $D\times L$ loop momenta
to a series of multiplications
of $D\times L$-dimensional matrices,
thus reducing greatly the number of numerical operations.

~\\
\noindent
{\bf 2.}
The method is most easily introduced by
treating an exactly solvable
class of integrals
\begin{eqnarray}
I^D_a=\int \frac{d^Dp}{(2\pi)^D}\frac{1}{(1+{\bf p^2})^a},
\label{@e1}\end{eqnarray}
whose integration via Schwinger's proper-time technique yields
\begin{eqnarray}
I^D_a=\frac{1}{(4\pi)^{D/2}}\frac{ \Gamma (a-D/2)}{ \Gamma (a)}.
\label{@e2}\end{eqnarray}
Our  perturbative evaluation
starts out by introducing
into (\ref{@e1}) a smallness parameter $ \epsilon $,
and considering the functions of $ \epsilon $
\begin{eqnarray}
I^D_a( \epsilon )=\int \frac{d^Dp}{(2\pi)^D}\frac{1}{(1+ \epsilon {\bf p^2})^{a/ \epsilon }}.
\label{@e3}\end{eqnarray}
For this function, we can easily find a perturbation expansion in powers of
$ \epsilon $, by rewriting it as follows:
\begin{eqnarray}
I^D_a( \epsilon )=\frac{1}{(4\pi a)^{D/2}}\int \frac{d^Dp}{(\pi/a)^{D/2}}
e^{-a{\bf p}^2-S_\epsilon({\bf p})},
\label{@ept}\end{eqnarray}
with
\begin{eqnarray}
 S_\epsilon({\bf p})  &\equiv& \frac{a}{ \epsilon }\log(1+ \epsilon {\bf p}^2)-a{\bf p}^2\\
&=&-a\left[
\frac{ \epsilon }{2}({\bf p}^2)^2
-\frac{ \epsilon^2 }{3}({\bf p}^2)^3
+\frac{ \epsilon^3 }{4}({\bf p}^2)^4+\dots
\right] .
\label{@S}\end{eqnarray}
Considering $a{\bf p}^2$ as a free ``action",
and $ S_\epsilon({\bf p}) $ as an ``interaction",
we
introduce the Gaussian expectation values
\begin{equation}
\langle f({\bf p})\rangle \equiv
\int \frac{d^Dp}{(\pi/a)^{D/2}}
e^{-a{\bf p}^2}f({\bf p}),
\label{@}\end{equation}
and rewrite the integral
(\ref{@ept}) as a harmonic expectation
\begin{eqnarray}
I^D_a( \epsilon )=\frac{1}{(4\pi a)^{D/2}}\langle
e^{-S_\epsilon({\bf p})}\rangle .
\label{@ept2}\end{eqnarray}
The exponential
$e^{- S_\epsilon({\bf p}) }$
 is now expanded
 in a Taylor series,
and the expectation values
$\langle S_\epsilon({\bf p})\cdots S_\epsilon({\bf p})\rangle $
are calculated via a Wick expansion
using the basic correlation function
\begin{equation}
\langle p_{i_1}p_{i_2}\rangle =G_{i_1i_2}=\frac{1}{2a} \delta _{i_1i_2},
~~~~i=1,\dots,D.
\label{@}\end{equation}
The nonzero expectation values of
higher products
\begin{equation}
\langle p_{i_1}\cdots p_{i_{2n}}\rangle =G_{i_1\dots i_{2n}}
\label{@}\end{equation}
are calculated from the recursion relation
\begin{eqnarray}
&&\!\!\!G_{i_1\dots i_{2n}}=
G_{i_1i_2} G_{i_3i_4\dots i_{2n-1} i_{2n}}
+G_{i_1i_3} G_{i_2 i_4\dots i_{2n}}
\nonumber \\
&&~~~~+\dots
+G_{i_1i_{2n-1}} G_{i_2 i_3\dots i_{2n}}
+G_{i_1i_{2n}} G_{i_2 i_3\dots i_{2n-1}}.
\label{@}\end{eqnarray}
For example, we find
{\footnotesize
\begin{eqnarray}
~~I_1^1( \epsilon )&=&\frac{1}{(4\pi)^{1/2}}
\left(1+\frac{3 \epsilon }{8}
+\frac{25 \epsilon ^2}{128}
+\frac{105 \epsilon ^3}{1024}
+\frac{1659 \epsilon ^4}{32768}
+\dots\right),\nonumber \\
I_2^3( \epsilon )&=&\frac{1}{(8\pi)^{3/2}}\left(1
\!+\!\frac{15 \epsilon }{16}
\!+\!\frac{385 \epsilon ^2}{512}
\!+\!\frac{4725 \epsilon ^3}{8192}
\!+\!\frac{228459 \epsilon ^4}{524288}+
\dots\right).\nonumber
\label{@}\end{eqnarray}}
~\\
The corresponding cumulant expansions are
\begin{eqnarray}
~~I_1^1( \epsilon )&=&\frac{1}{(4\pi)^{1/2}}e^{\frac{3 \epsilon }{8} \epsilon
+\frac{ 1}{8} \epsilon^2
+\frac{3 }{64} \epsilon^3
+\frac{ 1}{64}  \epsilon^4
+\dots},\label{@log0} \\
I_2^3( \epsilon )&=&\frac{1}{(8\pi)^{3/2}}e^{\frac{15  }{16} \epsilon
+\frac{5 }{16} \epsilon^2
+\frac{75}{512} \epsilon^3
+\frac{41 }{512} \epsilon^4
+\dots}.
\label{@log}\end{eqnarray}
Evaluating the exponents at $ \epsilon =1$ successively
as partial sums of orders 1,2,3,\dots,15
[not all terms are written down in
(\ref{@log0}) and (\ref{@log}),
for lack
of space], the relative deviations from the respective exact
values $1/2$ and $1/8\pi$ [which follow from Eq. (\ref{@e2})]
are shown in Table~\ref{@tab1} [columns marked with (ps)].
The deviation sequences tend slowly toward zero,
and grow again for larger order as a consequence
of the asymptotic nature of the expansions.

\begin{figure}[tbhp]
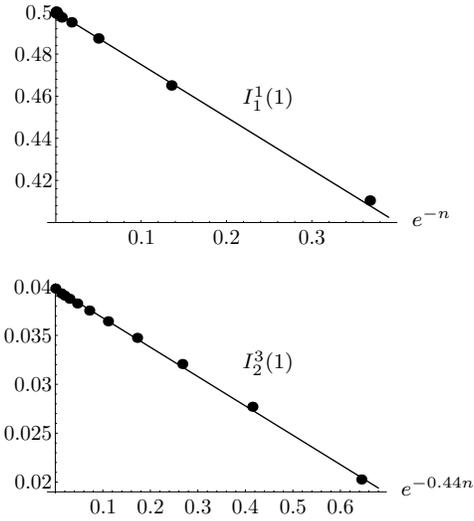

\input pl11.tps  \\
\caption[]{Exponentially fast convergence of the
first ten
approximations, obtained from
the
successive partial sums of the exponents in
(\ref{@log0}) and (\ref{@log}),
against their respective
exact values $1/2$ and $1/8\pi\approx 0.397884\dots~$. The number $n$ denotes
the order of the approximations. }
\label{fig0}\end{figure}

The partial sums yield integrals whose first ten terms
tend exponentially fast
toward the exact results, as shown in Figs.~\ref{fig0}.
The convergence is accelerated
by approximating the successive exponents
\`a la
 Pad\'e, leading rapidly to accurate results, as we can see in
the right-hand columns of
Table~\ref{@tab1}.

\begin{figure}[tbhp]
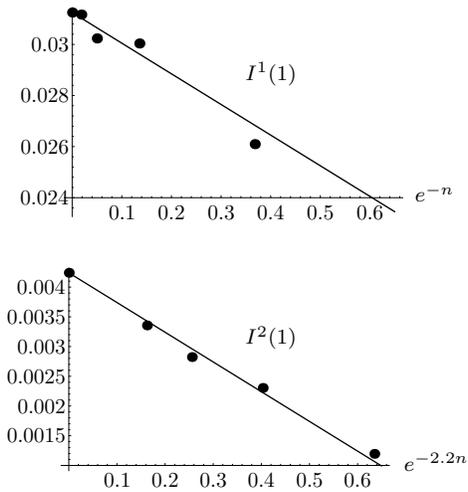

\input pl.tps  \\
\caption[]{Exponentially fast convergence of the
approximation
sequences
(\ref{@first}) and
(\ref{@second})
against their respective
exact values $1/32$ and $0.00424027\dots~$. The number $n$ denotes
the order of the approximations. }
\label{fig1}\end{figure}

~\\
\noindent
{\bf 3.}
As a nontrivial example
consider now the Feynman integral
\begin{eqnarray}
I^D&=&
\int \frac{d^Dp_1}{(2\pi)^D}
\int \frac{d^Dp_2}{(2\pi)^D}
\int \frac{d^Dp_3}{(2\pi)^D}\nonumber \\
&&\times
\frac{1}{1+{\bf p}_1^2}~
\frac{1}{1+{\bf p}_2^2} ~
\frac{1}{1+{\bf p}_3^2}~
\frac{1}{1+(
{\bf p}_1+
{\bf p}_2 +
{\bf p}_3)^2
}  .
\label{@12}\end{eqnarray}
We introduce again the
smallness parameter $ \epsilon $ as in (\ref{@e3}),
and rewrite the integral (\ref{@12})
as
\begin{eqnarray}
I^D( \epsilon )&=&
\int \frac{d^Dp_1}{(2\pi)^D}
\int \frac{d^Dp_2}{(2\pi)^D}
\int \frac{d^Dp_3}{(2\pi)^D}\nonumber \\&&\times e^{-\left[
{\bf p}_1^2+{\bf p}_2^2+{\bf p}_3^2
+(
{\bf p}_1+
{\bf p}_2 +
{\bf p}_3)^2
\right] -S_\epsilon({\bf p}_1,{\bf p}_2,{\bf p}_3)},
\label{@13}\end{eqnarray}
with a free ``action"
$
{\bf p}_1^2+{\bf p}_2^2+{\bf p}_3^2
+(
{\bf p}_1+
{\bf p}_2 +
{\bf p}_3)^2
$,
and an ``interaction"
\begin{eqnarray}
 &&\!\!\!\!\!\!\!\!S_\epsilon({\bf p}_1,{\bf p}_2,{\bf p}_3) =
 \sum_{a=1}^3
\left[ \frac{1}{ \epsilon }\log(1+ \epsilon {\bf p}_a^2)- {\bf p}_a^2\right]
\nonumber \\
&&
~~~~~~+\frac{1}{ \epsilon }\log\left[1+ \epsilon
\left(\sum_{a=1}^3{\bf p}_a\right)^2 \right]
-\left(\sum_{a=1}^3{\bf p}_a\right)^2
.
\label{@}\end{eqnarray}
The free action is now rewritten as
a quadratic form
$\frac{1}{2}p_{ai}K_{ai,bj}
p_{bj}$,
with the matrix
\begin{eqnarray}
K_{ai,bj}=k_{ab} \delta _{ij},~~~k_{ab}=
2\left(
\begin{array}{lll}
2&1&1\\
1&2&1\\
1&1&2
\end{array}
\right),~~~a=1,2,3.
\nonumber
\label{@}\end{eqnarray}
where $ \delta _{ij}$ is the $D$-dimensional Kronecker symbol.
We now define harmonic expectation values
\begin{eqnarray}
&&\langle f({\bf p}_1,{\bf p}_2,{\bf p}_3)\rangle\equiv (2\pi)^{3D/2}(\det K)^{1/2}
\label{@expv} \\&&\times \!
\int \!\frac{d^Dp_1}{(2\pi)^D} \!
\int\! \frac{d^Dp_2}{(2\pi)^D} \!
\int\! \frac{d^Dp_3}{(2\pi)^D}
f({\bf p}_1,{\bf p}_2,{\bf p}_3) e^{-\frac{1}{2}p_{ai}K_{ai,bj}
p_{bj}}  \nonumber
\label{@}\end{eqnarray}
where
\begin{eqnarray}
\det K= \det k^D=32^D.\nonumber
\label{@}\end{eqnarray}
The basic correlation functions are
\begin{eqnarray}
\langle p_{ai}p_{bj}\rangle &=&
K^{-1}_{ai,bj}=
G_{ai,bj}=
g_{ab} \delta _{ij},\nonumber \\
g_{ab}&=&k^{-1}_{ab}=\frac{1}{8}
\left(
\begin{array}{rrr}
3&-1&-1\\
-1&3&-1\\
-1&-1&3
\end{array}
\right).\nonumber
\label{@}\end{eqnarray}
In terms of (\ref{@expv}),
the integral (\ref{@13})
can now be considered as a harmonic expectation value
\begin{eqnarray}
I^D( \epsilon )={(2\pi)^{-3D/2}}(\det K)^{-1/2}
\langle e^{ -S_\epsilon({\bf p}_1,{\bf p}_2,{\bf p}_3)}\rangle,
\label{@16}\end{eqnarray}
which is evaluated perturbatively
by
expanding
the exponential
into a power series,
and forming all Wick contractions.
The results are for $D=1$ and 2:
\begin{eqnarray}
&&I^1( \epsilon )= \frac{1}{(2\pi)^{3/2}}\frac{1}{ \sqrt{32}}
e^{\frac{27}{32}\epsilon + \frac{9}{64}\epsilon^2 +
\frac{24}{4096}\epsilon^3 +
  {\frac{249}{8192}}\epsilon^4
+\dots},\nonumber \\
&&I^2( \epsilon )= \frac{1}{(2\pi)^3}\frac{1}{ \sqrt{32^2}}
e^{{\frac{9}{4}}\epsilon + {\frac{21}{32}}\epsilon^2
+ {\frac{13}{64}}\epsilon^3 +
  {\frac{177}{1024}}\epsilon^4
+\dots},
\label{@}\end{eqnarray}
and the successive evaluation of the exponent
at $ \epsilon =1$ yields the approximations
\begin{eqnarray}
&&0.026097,~\,0.030038,~\,0.030236,~\,0.031169,
~
\label{@first}\\
&&0.001195,~\,0.002304,~\,0.002823,~\,0.003356,
\label{@second}\end{eqnarray}
which converge against the
precise values of the Feynman integral $1/32=0.03125$ and
$0.00424027\dots~$, respectively.
These tend
exponentially fast
against the exact value,
as seen in Fig.~\ref{fig1}.

The same Pad\'e  approximants as before yield the
numbers
\begin{eqnarray}
&&0.030894,~~~~~~~0.030171,~~~~~~~~\,0.03064,\\
&&0.003019,~~~~~~~0.003091,~~~~~~~~\,0.00309,
\label{@}\end{eqnarray}
which, surprisingly,
show no improvement with respect
to the
sequences
(\ref{@first}),
(\ref{@second}).

~\\
\noindent
{\bf 4.}
So far we have calculated only vacuum diagrams.
There is no problem in extending the method to diagrams
with external legs.
The first nontrivial Feynman integral
in the self-energy of a $ \phi^4$-theory, for example,
is
\begin{eqnarray}
I^D({\bf q})&=&
\int \frac{d^Dp_1}{(2\pi)^D}
\int \frac{d^Dp_2}{(2\pi)^D}
\nonumber \\
&&\times
\frac{1}{1+({\bf p}_1+{\bf q})^2}~
\frac{1}{1+{\bf p}_2^2} ~
\frac{1}{1+(
{\bf p}_1+
{\bf p}_2)^2
}  .
\label{@12a}\end{eqnarray}
which we rewrite as
\begin{eqnarray}
\!\!\!\!\!\!\!\!\!\!
I^D( {\bf q};\epsilon )&=&   e^{-{\bf q}^2}
\int \frac{d^Dp_1}{(2\pi)^D}
\int \frac{d^Dp_2}{(2\pi)^D}
\nonumber \\&&\times e^{-\left[
{\bf p}_1^2+
{\bf p}_2^2
+(
{\bf p}_1+
{\bf p}_2)^2
\right] -S_\epsilon({\bf p}_1,{\bf p}_2,{\bf q})-2{\bf p}_1{\bf q}},
\label{@13e}\end{eqnarray}
with a free ``action"
$
{\bf p}_1^2+{\bf p}_2^2
+(
{\bf p}_1+
{\bf p}_2)^2
$,
a ${\bf q}$-dependent ``interaction"
\begin{eqnarray}
 &&\!\!\!\!\!\!\!\!S_\epsilon({\bf p}_1,{\bf p}_2,{\bf q}) =
 \sum_{a=1}^3
\left[ \frac{1}{ \epsilon }\log(1+ \epsilon {\bf p}_a^2)- {\bf p}_a^2\right]
\nonumber \\
&&
~~~~~~+\frac{1}{ \epsilon }\log\left[1+ \epsilon
\left(\sum_{a=1}^3{\bf p}_a\right)^2 \right]
-\left(\sum_{a=1}^3{\bf p}_a\right)^2
,
\label{@}\end{eqnarray}
where we have set ${\bf p}_3\equiv{\bf p}_1+{\bf p}_2 $.
After removing the
extra source term
$2{\bf p}_1{\bf q}$
by a shift of the integration variable
${\bf p}_1\rightarrow
{\bf p}_1-{\bf q}$,
which produces an exponential  factor $e^{2 k_{11}{q}^2}$,
the integral can be treated as before,
yielding an expansion in powers of $ \epsilon $ with ${\bf q}^2$-dependent coefficients.

~\\
\noindent
{\bf 5.}
The above treatment of Feynman diagrams
offers the interesting possibility of
deriving a completely new
type of power series expansions
for the critical exponents
of $\phi^4$-theories
in three dimensions, in which
expansion coefficients
are
calculable {\em exactly\/}, in contrast to the known expansions of Ref.~\cite{MN}.
For this we observe that
our power series for the Feynman integrals
can be reexpanded in powers
of $g$ by inserting $ \epsilon \equiv g/(m+g)$.
Then
 $I_1^1( \epsilon )$ and
 $I_2^3( \epsilon )$ become
\begin{eqnarray}
I_1^1(\bar g )&=&\frac{1}{(4\pi)^{1/2}}\exp\left(\frac{3 \bar g }{8}
\!-\!\frac{ \bar g^2}{4}
\!+\!\frac{11 \bar g ^3}{64}
\!-\!\frac{ \bar g ^4}{8}
\!+\!\dots\right),\nonumber \\
I_2^3( \bar g )&=&\frac{1}{(4\pi)^{3/2}}\exp\left(\frac{15 \bar g }{16}
\!-\!\frac{5 \bar g ^2}{8}
\!+\!\frac{235\bar g ^3}{512}
\!-\!\frac{23\bar g ^3}{64}
\!+\!\dots\right),\nonumber
\label{@log}\end{eqnarray}
where $\bar g\equiv g/m$.
The coefficients  have now alternating signs.
These expansions must be evaluated in the strong-coupling limit
$\bar g\rightarrow \infty$.
This is again possible with the help of diagonal
Pad\'e approximations, and the results
are precisely the same as those in Table~\ref{@tab1}.
Note that the choice of the parameter $m$ is irrelevant in this limit.

Now we recall that  the critical behavior of $\phi^4$-theory can also be
derived from  power series expansions
of the renormalization constants in the strong-coupling limit
$g_B/m_B\rightarrow \infty$, where $g_B$ and $m_B$ are the bare coupling strength and
mass of the theory.
Let us identify the above expansion parameter
$\bar g$ with
 $c g_B/m_B$, where $c$ is an arbitrary parameter
which may be chosen at the end to optimize the speed
of convergence.
Then we insert the
expansions of
the Feynman integrals
in powers of $g_B/m_B$
into the Feynman
diagrams of the perturbation
expansions of the renormalization constants
of the $\phi^4$-theory. This leads
to a new type of  series
in powers of $g_B/m_B$ whose coefficients
are determined exactly.
By going to the strong-coupling limit $g_B/m_B\rightarrow \infty$
as described in Ref.~\cite{sc,seven},
we can obtain the
critical exponents in three dimensions
in a completely new way.

From the connected vacuum diagrams we obtain an expansion
of the form
\begin{equation}
W=m_B^3Z=m_B^3\left(1+a_1\bar g_B+a_2
\bar g_B^2+\dots\right) .
\label{@Z}\end{equation}
For a comparison with experiment
one assumes the bare mass to go linearly to zero
as the temperature $T$ approaches the
critical temperature, i.e. $m_B^2\propto(T/T_c-1)$.
The vacuum energy $W$ has then for small
$m_B$ the scaling behavior
$(T/T_c-1)^{2- \alpha }\propto m_B^{4-2 \alpha }$, where $ \alpha =2-3 \nu $
is the critical exponent
of the  peak in the specific heat in three dimensions ($ \nu $is the exponent of the correlation length).
Since $\bar g_B $ grows for small $m_B$
like $1/m_B$, the power series for $Z$ in (\ref{@Z}) must
behave like $Z\propto  \bar g_B^{2 \alpha -1}$.
The critical exponent $ \alpha $ is therefore found from
the strong-coupling limit  of the power series
\begin{equation}
 \alpha(\bar g_B) =\frac{1}{2}(1+d \log Z/d \log \bar g_B),
\label{@al}\end{equation}
evaluated as explained in Ref.~\cite{sc,seven}.

Also the exponent $ \omega $ governing
the approach to scaling can be obtained from the series
(\ref{@al}), since  $\alpha (\bar g_B)$ approaches
$ \alpha $ for large $\bar g_B$
like
$\alpha+$const$/\bar g_B^ \omega $.
Thus $ \omega $ is found from the strong-coupling limit
of the series $ \omega(\bar g_B) =-1  - d \log \alpha'(\bar g_B)/d \log \bar g_B\alpha'(\bar g_B)$
(see \cite{nobeta} for more details).

The third independent critical exponent may be derived from the
perturbative corrections to the
field normalization.
For this we take the expansion for the self-energy $ \Sigma ({\bf q})$
and extract the coefficient of the ${\bf q}^2$ term.
This is done most simply by
differentiating  all Feynman integrals of the two-point function
using the rule (in $D$ dimensions)
\begin{equation}
\frac{\partial }{\partial {\bf q}^2}=\frac{1}{2D}
\frac{\partial ^2}{\partial q_\mu\partial q_\mu},
\label{@}\end{equation}
and setting ${\bf q}=0$.
The momentum ${\bf q}$ always flows through
one or more lines of the diagram, allowing the
differentiations
to be performed as follows:
\begin{eqnarray}
   \frac{\partial ^2}{\partial q_\mu\partial q_\mu} \frac{1}{({\bf p}-{\bf q})^2+m^2}
    &=&\frac{-2D}{[({\bf p}-{\bf q})^2+m^2]^2}\nonumber \\
&&+\frac{8\,({\bf p}-{\bf q})^2}
{[({\bf p}-{\bf q})^2+m^2]^3}
     .
\label{@}\end{eqnarray}
After setting ${\bf q}=0$ in the resulting Feynman integrals,
the calculation of the
field normalization  proceeds using the above formulas
for the vacuum diagrams.

Note that
the $ \epsilon $-dependent Feynman integrals are all finite
for small $ \epsilon $.
However, as
$ \epsilon $
 increasing towards unity,
a pole
is encountered
 at $ \epsilon =2/3$.
Thus, the expansion of a pole term
$\propto 1/( \epsilon -2/3)$
should be subtracted from
the $ \epsilon $-expansions of divergent integrals
before one can evaluate
perturbation series at
$ \epsilon =1$, i.e., before going to the strong-coupling limit
of $\bar g_B$.

~\\
\noindent
{\bf 6.}
Summarizing we have shown that just as the generation
of all Feynman diagrams \cite{rec}, also their evaluation
can now be automatized by computer algebra, with the additional
advantage that all expansion coefficients emerge as exact numbers.

The actual calculation
of diagrams with eight and
and more loops will, of course, require a considerable amount of work.
It remains to be seen whether the new perturbation expansions
converge more rapidly against the exact results than the presently known
purely numerical ones.

%
%

%
\begin{table}[tbhp]
\caption[]{Relative deviations of approximations
$I_1^1( 1 )$ and
$I_2^3( 1 )$
 from their exact values, once from successive partial sums (ps),
once from successive Pad\'e approximations.
}
\begin{tabular}{r|rr||l|rrrrrr}
$n$&$I_1^1(1)$~ps&$I_2^3(1)$~ps&[N,M]~~~&$I_1^1(1)$~Pad\'e&$I_2^3(1)$~Pad\'e\\
\hline
1&   0.17910&0.49063&[0,1] ~~&-0.391430  &-3.85230\phantom{$1^{-5}$}\\
2&   0.06980&0.30377&[1,1] ~~& 0.009816  & 0.18603\phantom{$1^{-5}$}\\
3&   0.02516&0.19394&[1,2] ~~& 0.002101  & 0.09851\phantom{$1^{-5}$}\\
4&   0.00981&0.12674&[2,2] ~~& 0.000935& 0.02490\phantom{$1^{-5}$}\\
5&   0.00516&0.08416&[3,3] ~~& 0.001821& 0.00309\phantom{$1^{-5}$}\\
6&   0.00256&0.05655&[4,4] ~~& 0.000637&-0.00062\phantom{$1^{-5}$}\\
7&   0.00027&0.03836&[5,5] ~~& 0.000652&-0.00062\phantom{$1^{-5}$}\\
8&  -0.00021&0.02625&[6,6] ~~& 0.000325& 4.8349~$10^{-5}$\\
9&   0.00124&0.01810&[7,7] ~~& 0.000325& 8.0048~$10^{-6}$\\
10&  0.00114&0.01255&[8,8] ~~& 0.000192& 3.3873~$10^{-5}$\\
11& -0.00273&0.00875&[9,9] ~~& 0.000194  &1.6181~$10^{-5}$\\
12& -0.00275&0.00613&[10,10]&0.000127  &1.8694~$10^{-5}$\\
13&  0.01001&0.00432&[11,11]&0.000130  &1.0960~$10^{-5}$\\
14&  0.01000&0.00305&[12,12]&0.000090  &1.1146~$10^{-5}$\\
15& -0.05026&0.00216&[13,13]&0.000095  &7.3635~$10^{-6}$\\
\hline \\
\end{tabular}
\label{@tab1}\end{table}

\end{document}